\documentclass{elsart}
\usepackage{amssymb}
\usepackage{graphicx,color}

\begin {document}

\begin{frontmatter}
\title{Success of collinear expansion in the calculation of
induced gluon emission}

\author{Xin-Nian Wang}
\address{Nuclear Science Division, MS 70R0319,
Lawrence Berkeley National Laboratory, Berkeley, California 94720}

\begin{abstract}
We clarify the confusion in a recent paper by Aurenche, Zakharov and
Zaraket (AZZ) \cite{azz} over the procedure and region of validity 
of the collinear approximation in the twist expansion approach to
induced gluon emission in deeply inelastic scattering (DIS) off
a nucleus target. We point out that, in this approach to the semi-inclusive 
spectrum, the transverse momentum $\vec\ell_\perp$ of the induced 
gluon must be fixed in the collinear expansion in the 
transverse momentum $\vec k_\perp$ of the initial partons, therefore
the result is valid for $\langle k_\perp^2\rangle \ll 
\ell_\perp^2 \ll Q^2$. In the twist-four contribution, one can 
single out the double-hard term corresponding to collinear quark-gluon 
Compton scattering which can be calculated independently of the 
collinear expansion approach. We will discuss the connection between 
the collinear approximation in the twist expansion approach and the 
small $k_T$ approximation of the results in the Light-Cone Path 
Integral (LCPI) and Gyulassy-Levai-Vitev (GLV) opacity expansion 
approach. We point out the misconstrued variable 
change by AZZ before the $k_T$ expansion in LCPI and opacity expansion
approach, without which one obtains the same result for the induced 
gluon spectrum under collinear approximation as in the twist expansion
approach. We also show that corrections beyond the collinear approximation
to the transverse momentum integrated gluon spectrum within the static
potential model in the GLV approach give rise to a logarithmic factor 
difference from the result of collinear approximation.

\end{abstract}


\end{frontmatter}

\section{Introduction}

Jet quenching due to parton energy loss in dense medium provides an
excellent probe of the hot quark matter produced in high-energy
heavy-ion collisions \cite{wg90}. Such predicted phenomenon was
indeed observed in experiments at the Relativistic Heavy-ion
Collider (RHIC) \cite{phenix,star}. Phenomenological studies of
the observed jet quenching depend crucially on our theoretical
understanding of parton propagation and energy loss in dense
medium which is dominated by induced gluon radiation via multiple
scattering. There have been, therefore, a plethora of theoretical studies
of induced gluon emission from a propagating
parton \cite{Gyulassy:1993hr,Baier:1996sk,Zakharov:1996fv,Wiedemann:2000za,
Gyulassy:2000er,guow}. One
approach to the study of induced gluon emission is based on higher-twist
expansion of multiple parton scattering cross section 
in a nuclear medium in the framework
of collinear factorization approximation \cite{guow,zw} which can also be
applied to parton propagation in hot medium. The advantage of higher-twist
approach is the natural formulation of the parton propagation problem in
terms of medium modified parton fragmentation function which is the only
physical observable of the jet quenching as a result of induced gluon
emission and parton energy loss.

The higher-twist approach to the problem of induced gluon emission is so far
limited to twist-four contributions in the twist expansion. This is equivalent
to the leading order approximation in the opacity expansion approach by Gyulassy,
Levai and Vitev (GLV) \cite{Gyulassy:2000er}.
The induced gluon spectra from these two approaches can
be shown to be equivalent under the twist and opacity expansion
approximations \cite{tmg}. However, Aurenche, Zakharov and Zaraket (AZZ)
claim in a recent note \cite{azz} that the higher-twist approach
``fails'' to produce the correct gluon spectra as in the
Light-cone Path Integral (LCPI) and opacity expansion approach by
Gyulassy, Levai and Vitev (GLV). We want to demonstrate that such an
unjustified claim comes from their confusion over the procedure and
validity of the collinear expansion in the higher-twist approach.
A misconstrued variable change in LCPI and GLV results also lead
to their conclusion on the collinear approximation in the $k_T$
factorized formulation. We demonstrate that these results under the
same approximation are equivalent to each other.

For semi-inclusive cross section of induced gluon emission, the collinear
expansion in terms of the initial parton transverse momentum $k_T$ 
must be made for a fixed value $\langle k_T^2\rangle \ll \ell_T^2 \ll Q^2 $
of the gluon's transverse momentum. For the total parton energy loss
which involves integration over the gluon's transverse momentum, we will
argue that the higher-twist expansion is a good approximation for
large initial jet energy $q^-\gg R_A \langle k_T^2\rangle$, where $R_A$ is the
nuclear size.

\begin{figure}
\begin{center}
\includegraphics[width=1.7in,height=1.in]{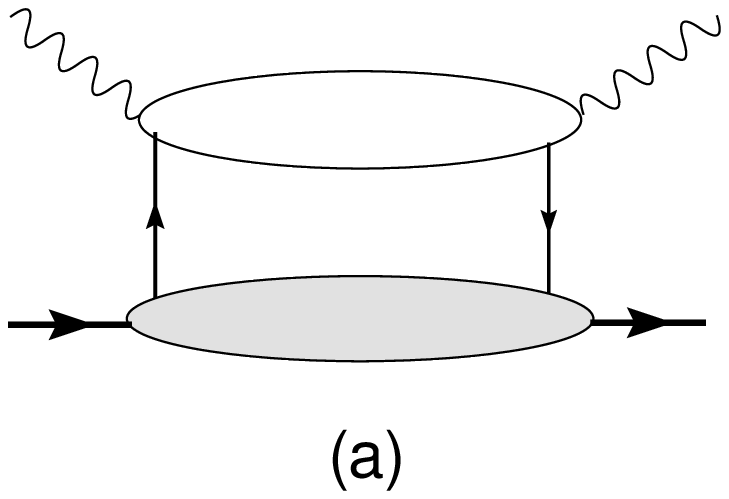}
\includegraphics[width=1.7in,height=1.in]{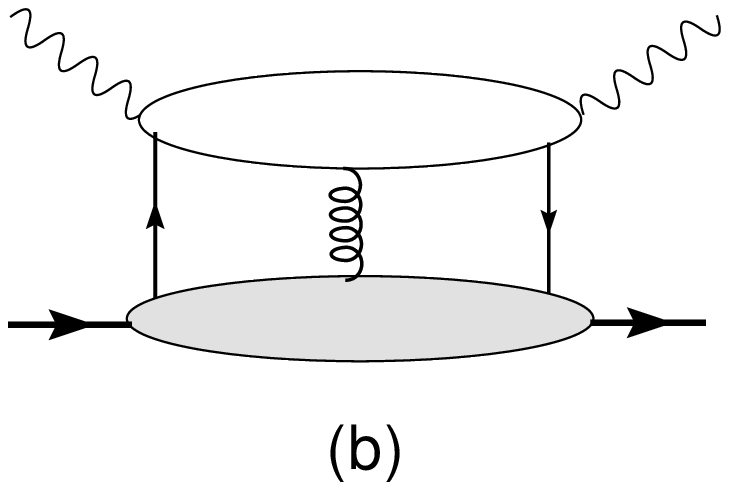}
\includegraphics[width=1.7in,height=1.in]{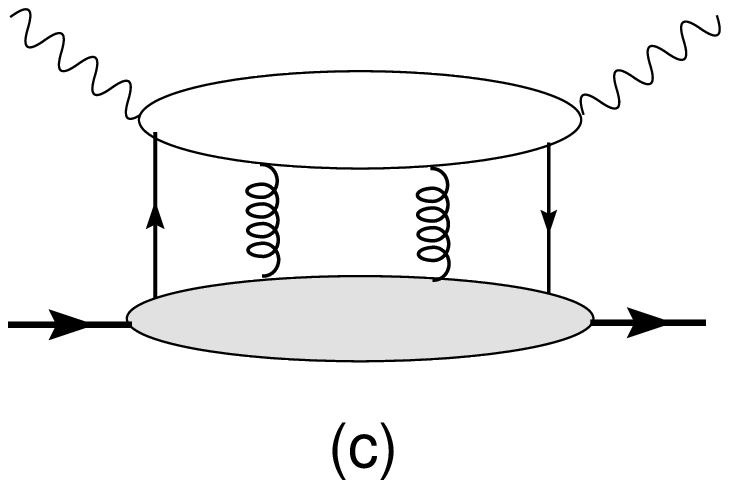}
\end{center}
\caption{Diagrams of DIS with multiple gluon interaction}
\label{fig1}
\end{figure}

\section{Collinear factorization and twist expansion}

In the twist expansion approach to multiple parton scattering, one considers
interaction between a fast parton and the target nucleus (or hot medium) via
exchange of soft gluons as shown in Fig.~\ref{fig1} for the case of deeply
inelastic scattering (DIS) off a large nucleus. In the framework of
collinear factorization, one normally chooses a covariant gauge and
considers $A^+$ as the largest component of the gauge field.
Since the longitudinal momentum per target
nucleon $p=[p^+,0,\vec{0}_\perp]$ is the largest
momentum component in the process, the twist expansion
procedure involves expanding the hard part $H_n(k_T)$ of the
quark-$n$-gluon interaction in the transverse momentum $k_T$ of
the initial gluon fields and the transversely
polarized gluon fields $A_\perp$. Using generalized Ward
identities one can relate the derivatives of the hard parts
$\partial_{k_T}^n H(k_T)_{k_T=0}$ of quark-longitudinal
-gluon interaction and the collinear hard parts of
quark-transverse-gluon interaction, and combine them to
produce gauge invariant higher-twist contributions to
the DIS cross section (see Ref.~\cite{lw} for a detailed illustration).
After integration over the
transverse momentum of initial gluon fields, the final results
are given by the convolution of collinear parton scattering
cross sections and transverse momentum integrated parton
distribution or correlation matrix elements.  The leading
twist-four and nuclear enhanced contributions to the hadronic
tensor of DIS off a nucleus involve two longitudinal 
gluon fields \cite{note1} as shown in Fig.~\ref{fig1}c and
can be expressed as
\begin{eqnarray}
W_2 &=&
\int \frac{dy^-}{2\pi} dy_1^- dy_2^-
\frac{d^2\xi_T}{(2\pi)^2}
d^2k_T e^{-i\vec k_T\cdot\vec\xi_T} H_2^{--}(k_T)
 \nonumber\\
&&\hspace{.5in}\times \langle A \mid
\bar{\psi}(0) \frac{\gamma^+}{2}
A^+(y_1^-,0_T) A^+(y_2^-,\xi_T)\psi(y^-) \mid A\rangle,
\end{eqnarray}
where we have suppressed the Lorentz indices of the
electromagnetic currents and other kinematic variables in
the perturbative hard part $H_2^{--}(k_T)$ of the multiple parton 
scattering with longitudinal gluon fields.
 Summations over color indices of the field
operators in the matrix
and average over the color indices of the initial state partons
in the hard part are understood. In the collinear factorization scheme,
one makes a collinear expansion of the hard part
\begin{equation}
H_2^{--}(k_T)=H_2^{--}(0)+k_T^i \frac{H_2^{--}(k_T)}{\partial k_T^i}\mid_{k_T=0}
+\frac{k_T^ik_T^j}{2}\frac{\partial^2 H_2^{--}(k_T)}{\partial k_T^i\partial k_T^j}
\mid_{k_T=0} +\cdots .
\label{eq:expand}
\end{equation}
In order to clarify the confusion in Ref.~\cite{azz} over the collinear
expansion, it is important to emphasize here that the longitudinal
gauge field $A^+$ is not a physical gluon field. Therefore, the
hard part $H_2^{--}(k_T)$ does not correspond to quark interaction with
physical gluons. This is apparent in the fact that the collinear
term in the above expansion does not vanish and is actually related
to the hard part with no longitudinal gluon interaction (Fig.~\ref{fig1}a),
$H_2^{--}(0)=(-ig)^2H_0$. One can prove in general this is true for quark
interaction with any number of longitudinal gluons.
After integration over the initial parton transverse momentum, which is
another important part of the collinear factorization scheme,
their contributions to the semi-inclusive cross section take the form
\begin{eqnarray}
&&H_0 \langle A\mid \bar\psi(0) \gamma^+\left[1-ig\int_0^{y^-}dy_1^-
A^+(y_1^-) \right.
\nonumber \\
&&\hspace{0.5in}+\left.(-ig)^2\int_0^{y^-}dy_1^-\int_0^{y_1^-}dy_2^-
A^+(y_1^-)A^+(y_2^-)+\cdots \right] \psi(y^-)\mid A\rangle ,
\end{eqnarray}
which becomes part of the leading twist contribution as the gauge link
in the gauge invariant quark distribution function
\begin{equation}
f_A^q(x)=\frac{1}{2}\int dy^- e^{ixp^+y^-}
\langle A\mid \bar\psi(0)\gamma^+{\rm P}e^{-ig\int_0^{y^-}d\xi^- A^+(\xi^-)}
\psi(y^-)\mid A\rangle \; .
\end{equation}
The interaction between a propagating quark and soft longitudinal
gauge fields will only result in an eikonal line along the light-cone.
This does not correspond to any physical scattering because one can get
rid of it by choosing a proper (physical) gauge. This is the
basic idea behind the proof of collinear factorization of
the leading twist cross section of DIS and Drell-Yan processes by
Collins, Soper and Sterman \cite{Collins:1985ue}.

The contribution from the linear term in Eq.~(\ref{eq:expand}) of the
collinear expansion vanishes for unpolarized targets. For the quadratic
term of the expansion one can combine $k_T^2$ with the longitudinal gauge
fields $A^+ A^+$ and obtain a quark-gluon correlation distribution
after partial integration over $k_T$,
\begin{eqnarray}
& & \int \frac{d^2\xi_T}{(2\pi)^2} \frac{dy^-}{2\pi}dy_1^-dy_2^- d^2k_T
e^{ixp^+y^- + ix_gp^+(y_1^- - y_2^-) -i\vec k_T\cdot\vec\xi_T} \nonumber \\
&& \hspace{1in} \times \langle A|\bar\psi(0) \frac{\gamma^+}{2}
k_T^2 A^+(y_1^-,0_T) A^+(y_2^-,\xi_T)\psi(y^-)
|A\rangle \nonumber \\
&\approx&  \int \frac{dy^-}{2\pi}
dy_1^-dy_2^- e^{ixp^+y^- + ix_gp^+(y_1^- - y_2^-)}
\langle A|\bar\psi(0) \frac{\gamma^+}{2} F^{+i}(y_1^-) F^+_i(y_2^-)\psi(y^-)
|A\rangle
\nonumber \\
&\approx& \pi \int dy_N^-\rho_A(y_N) f_A^q(x) x_gG_N(x_g),
\end{eqnarray}
where a factorized form of the quark-gluon
correlation is assumed, $\rho_A(y_N)$ is the nucleon density distribution
and $G_N(x_g)$ the gluon distribution function per nucleon inside
the nucleus. The momentum fraction $x_g$ carried by the initial gluon is
determined by the kinematics of each individual process.
Therefore, the leading twist-four contribution from the quadratic
term of the collinear expansion to the semi-inclusive DIS cross
section has a simple form
\begin{equation}
\pi \int dy_N^-\rho_A(y_N) \frac{1}{4}
\frac{\partial^2 H_2^{--}(k_T)}{\partial^2k_T}\mid_{k_T=0},
f_A^q(x) x_gG_N(x_g)
\label{crss0}
\end{equation}
which has a simple and intuitive interpretation of partonic
scattering between the fast quark and a {\it physical collinear} gluon
since the contribution is proportional to the $k_T$-integrated
gluon distribution
inside the nucleus. The quadratic derivative
term $\partial_{k_T}^2H_2^{--}(k_T)_{k_T=0}$ therefore corresponds to the
hard part of the actual physical collinear  quark-gluon scattering
cross section.

If one works in the physical gauge ($A^+=0$), the gauge link along the
light-cone
disappears (there will be transverse gauge link instead, see Ref.~\cite{ji}).
One therefore only has to consider quark interaction with the physical
(transverse) gluons. In this case, the hard part $H_2^{\perp\perp}(k_T)$, which
will be convoluted with the (physical) gluon distribution, corresponds to
partonic cross section of the physical quark-gluon scattering and is completely
different from the hard part $H_2^{--}(k_T)$ of quark interaction with
longitudinal gluons in the covariant gauge. One can in fact prove in general
the equivalence between the collinear hard part $H_2^{\perp\perp}(0)$ in
the physical gauge and the second derivative
$\partial_{k_T}^2H_2^{--}(k_T)_{k_T=0}$ of the hard part in the covariant
gauge \cite{lwz2}. We will illustrate this later in the case of induced
gluon radiation.

\section{Induced gluon spectra in higher-twist expansion}

One can calculate nuclear modification to the dijet cross section in 
DIS \cite{LQS2,LQS}, differential direct photon \cite{Guo:1995zk}
and Drell-Yan (DY) cross section \cite{Guo:1997it,Fries:1999jj} 
in $p+A$ collisions. The technique of higher-twist expansion has been
also been applied to calculate the induced gluon spectrum due to multiple
parton scattering in the DIS off a large nucleus or higher-twist contribution
to gluon radiative correction to the semi-inclusive DIS cross section 
in Refs.~\cite{guow,zw}. In this case, one should keep the final 
gluon transverse momentum $\ell_T$ {\it fixed} in the collinear expansion 
of the hard part in the initial parton transverse momentum $k_T$.
This is the origin of the flaw that leads to AZZ's conclusion in Ref.~\cite{azz}
about higher-twist approach to induced gluon spectrum.

\begin{figure}
\begin{center}
\includegraphics[width=3.5in,height=2.0in]{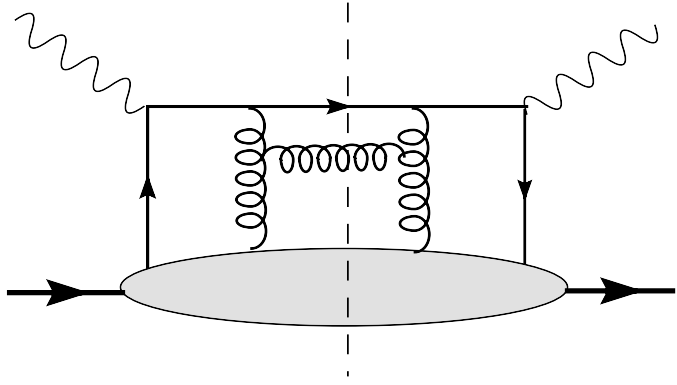}
\end{center}
\caption{Diagrams of induced gluon emission in DIS}
\label{fig2}
\end{figure}

The dominant contribution in the twist expansion approach mainly
comes from the process in Fig.~\ref{fig2}. The hard part of the
contribution to the semi-inclusive cross section before the collinear
expansion is (Eq.~(A.18) in Ref.~\cite{zw}),
\begin{eqnarray}
H_2^{--}(k_T)&=&
\frac{1}{(\vec{\ell_T}-\vec{k_T})^2}\, \frac{\alpha_s}{2\pi}\,
 C_A\frac{1+(1-z)^2}{z} \frac{2\pi\alpha_s}{N_c}
e^{ixp^+y^-+i(x_D+x_L)p^+(y_1^--y_2^-)}
\nonumber \\
&\times&[e^{i(x_D/z+x_L)p^+y_2^-}-1]
[e^{i(x_D/z+x_L)p^+(y^- - y_1^-)}-1],
\label{hard}
\end{eqnarray}
where
\begin{eqnarray}
x_L=\frac{\ell_T^2}{2p^+q^-z(1-z)}, \;\;\;\;
x_D=\frac{k_T^2-2\vec k_T\cdot\vec \ell_T}{2p^+q^-(1-z)}, \\
x_D/z+x_L=\frac{(\vec \ell_T-\vec k_T)^2}{2p^+q^-z(1-z)}.
\end{eqnarray}
Note that the overall phase factor quoted in Ref.~\cite{azz} is different
from $\exp[ixp^+y^-+i(x_D+x_L)p^+(y_1^--y_2^-)]$ of the actual result.
However, this does not affect the following argument.

First, one notices that the collinear limit of the hard part
$H_2^{--}(0)$ is finite and is the same as the vacuum gluon bremsstrahlung
except the phase factors. It combines with the collinear
limits of hard parts from other cut diagrams to form the gauge link
in the quark distribution function in the vacuum gluon (leading twist)
bremsstrahlung process in semi-inclusive DIS. To calculate higher-twist
contributions in the twist expansion approach, one keeps the
second order in the
collinear expansion of the above hard parts in $k_T$ for fixed
value of the gluon transverse momentum $\ell_T$. Therefore, one
needs only to evaluate the second derivative of the above hard
part, $\partial_{k_T}^2H_2^{--}$. The dominant contribution comes
from differentiating the factor $1/(\vec{\ell_T}-\vec{k_T})^2$.
As explained in the previous section, one combines the quadratic
term $k_T^2$ and the longitudinal gluon fields $A^+ A^+$ to form
the quark-gluon correlation function after partial integration.
The corresponding contribution to the gluon radiation spectrum is
\begin{eqnarray}
\frac{dN_2}{dzd\ell_T^2}=\frac{\alpha_s}{2\pi}\,
 C_A\frac{1+(1-z)^2}{z} \frac{1}{\ell_T^4}
\frac{2\pi\alpha_s}{N_c} \frac{T_{qg}^A(x,x_L)}{f_A^q(x)},
\label{diff}
\end{eqnarray}
where
\begin{eqnarray}
T_{qg}^A(x,x_L)&=& \int \frac{dy^{-}}{2\pi}\, dy_1^-dy_2^-
(1-e^{-ix_Lp^+y_2^-})(1-e^{-ix_Lp^+(y^--y_1^-)}) e^{i(x+x_L)p^+y^-}
\nonumber  \\
&&\hspace{-0.5in}\times\theta(-y_2^-)\theta(y^- - y_1^-)
\frac{1}{2}\langle A | \bar{\psi}_q(0)\,
\gamma^+\, F_{\sigma}^{\ +}(y_{2}^{-})\, F^{+\sigma}(y_1^{-})\,\psi_q(y^{-})
| A\rangle ,\label{Tqg}
\end{eqnarray}
is the quark-gluon correlation function of the nucleus and $f_A^q(x)$ is
the nuclear quark distribution function.

The leading contribution to the second-order
derivative  $\partial_{k_T}^2H_2^{--}$
from the phase factor in Eq.~(\ref{hard}) will be linear in
$(y_1^- -y_2^-)/q^-$ or $y^-/q^-$ which
in general are suppressed by a factor $\ell_T^2 r_N /q^-$ ($r_N$
is the nucleon size) or $\ell_T^2/x_Bp^+q^-$ ($x_B$ is fractional momentum
of the initial quark) relative to the above leading contribution
for large jet energy $q^-$. There are many other power-suppressed
terms like these from other diagrams. They can be neglected for
$\ell_T^2\ll Q^2$. 

Eq.~(\ref{diff}) has both hard-soft, double-hard scattering and
their interferences. The double hard scattering is characterized by the
finite momentum fraction $x_L$ carried by the initial gluon while
in hard-soft contributions the initial gluon carries zero fractional
momentum. This result is consistent with that in Ref~\cite{LQS} for nuclear
enhancement of jet photoproduction where they consider only hard-soft
and double hard scattering, but not their interferences which
is not important for large values of $\ell_T^2$.

The higher-twist result also has a simple and intuitive partonic
interpretation. One can assume the quark-gluon correlation has a
factorized form (see Sec. II of Ref.~\cite{jorge} for details),
\begin{eqnarray}
T_{qg}^A(x,x_L)&=& A \pi \int dy_N^- \rho_A(y_N)
\left[ f^q_N(x+x_L)[xG_N(x)]_{x=0}+ f^q_N(x) x_LG_N(x_L)\right]
\nonumber \\
&\times& [1-\cos(x_Lp^+y_N^-)],
\end{eqnarray}
where $f^q_N(x)$ is the quark and $G_N(x)$ the gluon distribution
per nucleon, $\rho_A(y_N)$ is the nucleon density at location $y_N$
inside the nucleus $A$.

The second term in the above factorized quark-gluon correlation
corresponds to contribution from double hard scattering in which
the initial gluon carries finite momentum fraction $x_Lp^+$.
The corresponding differential higher-twist gluon spectrum is then
\begin{eqnarray}
\frac{dN_{2}^{(H)}}{dzd\ell_T^2}&=&\int dy_N^-\rho_A(y_N) \frac{\pi \alpha_s^2}{\ell_T^4}\,
 \frac{C_A}{N_c}\frac{1+(1-z)^2}{z}x_LG_N(x_L)
[1-\cos(x_Lp^+y_N^-)]
\nonumber \\
&\equiv&\int dy_N^- \rho_A(y_N) \frac{d\sigma^N_{qg}}{dzd\ell_T^2}
[1-\cos(x_Lp^+y_N^-)]
\label{diff2}
\end{eqnarray}
which can be intuitively interpreted as the differential number
of quark-gluon scattering in a collinear factorized form as the
quark propagates inside the nucleus,
where $d\sigma^N_{qg}/dzd\ell_T^2$ is the collinear
quark-gluon cross section on a nucleon target.

One can derive the above double hard scattering contribution from
the simple collinear factorized parton model by noting that
Fig.~\ref{fig2} in this case is
just the quark-gluon Compton scattering process.
Considering a quark with momentum $q^-$ scattering with a gluon
that carries a fractional momentum $xp^+$,
$q (q)+g(xp)\rightarrow d(p^\prime)+g(\ell)$, the cross section
can be written as
\begin{eqnarray}
d\sigma_{ab} &=&\frac{g^4}{2\hat s}
|M|^2_{ab\to cd}(\hat t/\hat s,\hat u/\hat s)
\frac{d^3\ell}{(2\pi)^3 2\ell_0}
2\pi \delta[(xp+q-\ell)^2] \nonumber \\
&=&\frac{g^4}{(4\pi)^2}
|M|^2_{ab\to cd}(\hat t/\hat s,\hat u/\hat s)
\frac{\pi}{\hat{s}^2}
\frac{dz}{z(1-z)}d\ell_T^2 \,\, \delta\left(1-\frac{x_L}{x}\right),
\label{eq:cross-el}
\end{eqnarray}
where $q=[0,q^-,0]$ and $xp=[xp^+,0,0]$ are momenta of the initial partons
and
\begin{equation}
\ell=\left[\frac{\ell_T^2}{2zq^-},zq^-, \vec{\ell}_T\right]
\end{equation}
is the momentum of the final gluon. With the given kinematics,
the on-shell condition in the cross section can be recast as
\begin{eqnarray}
(xp+q-\ell)^2=2(1-z)xp^+q^-\left(1-\frac{x_L}{x}\right),\,\,\,\,\,
x_L=\frac{\ell_T^2}{2z(1-z)p^+q^-} .
\end{eqnarray}
The Mandelstam variables of the collision are,
\begin{eqnarray}
\hat s&=&(q+xp)^2=2xp^+q^-=\frac{\ell_T^2}{z(1-z)}, \,\,\,\,
\hat t=(\ell-xp)^2=-z\hat s,
\nonumber \\
\hat u&=&(\ell-q)^2=-(1-z)\frac{x_L}{x}\,\hat s=-(1-z)\hat s,
\label{eq:mand}
\end{eqnarray}
where the on-shell condition $x=x_L$ is used.

With Eq.~(\ref{eq:cross-el}) and gluon distribution functions
$G_N(x)$, one can obtain the quark-gluon scattering contribution to the
quark-nucleon cross section,
\begin{eqnarray}
d\sigma_{qg}^N&=&\int d\sigma_{qg}G_N(x) dx\nonumber \\
&=& x_LG_N(x_L)
|M|^2_{qg\to qg}(\hat t/\hat s,\hat u/\hat s)
\frac{\pi\alpha_s^2}{\hat{s}^2}
\frac{dz}{z(1-z)}d\ell_T^2
\label{eq:cross-el2}
\end{eqnarray}

Using the quark-gluon scattering matrix element
\begin{eqnarray}
|M|^2_{qg\to qg}&=&
\frac{C_A}{N_c}\frac{\hat{s}^2+\hat{u}^2}{\hat{t}^2}
-\frac{C_F}{N_c}\frac{\hat{s}^2+\hat{u}^2}{\hat{u}\hat{s}}
\nonumber \\
&=&\left[\frac{C_A}{N_c}\frac{1+(1-z)^2}{z^2}
+\frac{C_F}{N_c}\frac{(1+(1-z)^2)}{1-z}\right]\, ,
\end{eqnarray}
the quark-gluon cross section on a nucleon target is
\begin{eqnarray}
d\sigma_{qg}^N&=& x_LG_N(x_L)
\frac{\pi\alpha_s^2}{\ell_T^4}
\left[\frac{C_A}{N_c}(1-z)+\frac{C_F}{N_c}z^2\right]
\frac{1+(1-z)^2}{z}
dzd\ell_T^2 ,
\label{eq:cross-el3}
\end{eqnarray}
which is equivalent to the result in Eq.~(\ref{diff2}) in the soft
limit ($z\rightarrow 0$).

The divergent factor $1/\ell_T^4$ in the induced gluon
spectrum due to double hard scattering arises
because of the collinear approximation in which we neglected
the transverse momentum of the initial partons. This is
related to the neglect of the $k_T$ dependence of the
phase factors in the hard parts in the collinear expansions
and other higher-twist (larger than four) contributions.
These approximations are no longer valid at 
small values of $\ell_T^2\ll \langle k_T^2\rangle$
in the twist expansion approach, since many other
contributions and processes will become
important which are neglected in the above result.
We have argued \cite{guow} that contributions of
these neglected  terms could be approximated by substituting
$\ell_T^4\rightarrow 1/\ell_T^2(\ell_T^2+\mu^2)$
and $x_LG_N(x_L)\rightarrow (x_L+x_\mu)G_N(x_L+x_\mu)$
in the collinear result with $x_\mu=\mu^2/2p^+q^-$ and $\mu$ is
the average transverse momentum of the medium gluon.
However, the interference between double hard and soft rescattering 
processes suppresses the induced spectra for small $\ell_\perp^2 R_A/2q^- \ll 1$.
Therefore, the final result in the collinear expansion will be a good 
approximation and insensitive to the regularization for large initial
quark energy $q^- \gg \hat R_A\langle k_T^2\rangle$. For corrections
beyond the twist-four contribution, one has to consider the nuclear
broadening of the transverse momentum 
$\langle k_T^2\rangle \sim R_A \hat q$ ($\hat q$ is the 
jet transport parameter \cite{jorge})

The double hard scattering process corresponds to elastic scattering
in which there is a finite energy transfer ($x_L$) from the medium gluon.
With the leading order contribution to the medium gluon 
distribution $xG_N(x)\sim \delta (x-1)$,
the corresponding energy loss can be proved to be the same as the
elastic energy loss \cite{elastic}. This is a unique feature of the twist 
expansion approach that is not included in all other approaches (BDMPS and GLV). 
On the other hand, the quantum non-locality of the quark-photon interaction 
in DIS has never been considered as an important feature of the twist expansion
approach.

One can similarly interpret the first term in Eq.~(\ref{Tqg})
which corresponds to the hard-soft process.  Here the gluon
radiation is induced by the hard photon-quark scattering and
subsequently the radiated gluon interacts with a soft gluon from another
nucleon with distribution $x G_N(x)_{x=0}$. One can
compare this part of gluon spectrum
\begin{eqnarray}
\frac{dN_2^{(S)}}{dzd\ell_T^2}&=&\int dy_N^-\rho_A(y_N)
\frac{\pi \alpha_s^2}{\ell_T^4}
 \frac{C_A}{N_c}\frac{1+(1-z)^2}{z}xG_N(x)_{x\approx 0}
\nonumber \\
&&\hspace{.5in}\times [1-\cos(x_Lp^+y_N^-)]
\label{diff3}
\end{eqnarray}
to the induced gluon spectra in LCPI and GLV approaches, especially
after substitution $xG_N(x)_{x\approx 0}\rightarrow x_\mu G_N(x_\mu)$
and $\ell_T^4\rightarrow 1/\ell_T^2(\ell_T^2+\mu^2)$ when the effect of
the finite transverse momentum of the initial gluon is considered.

Note that with higher order contributions to the medium gluon
distribution function $xG_N(x)$, the semi-inclusive spectrum from the double hard 
scattering is similar to the hard-soft scattering with a correction
on the order of $(\ell_T^2/Q^2)[x\partial_x G_N(x)]_{x\approx 0}$ which
can be neglected for small values of $\ell_T^2\ll Q^2$.

\section{Induced gluon emission in $k_T$-factorized form}

The differential spectrum for induced gluon emission via interaction
between the fast quark and the medium or initial gluon in the LCPI
formulation is obtained in the $k_T$ factorized form,
\begin{eqnarray}
\frac{dN_{\rm LCPI}}{dzd\ell_T^2}&=&\frac{1+(1-z)^2}{z}
\int dy_N^- \rho_A(y_N)
\int d^2k_T \frac{xdG_N(k_T^2,x)}{d\ln k_T^2} \widetilde H(\vec k_T),
\nonumber \\
\widetilde H(\vec k_T) &=&2\pi\alpha_s^2
\frac{\vec k_T\cdot\vec\ell_T}{\ell_T^2(\vec k_T-\vec\ell_T)^2}
\left[1-\cos\frac{(\vec\ell_T-\vec k_T)^2y_N^-}{2q^-z(1-z)}\right],
\label{zak}
\end{eqnarray}
where $q^-$ is the energy of the fast quark, $k_T$ the transverse
momentum of the medium or initial gluon, $\ell_T$ the transverse
momentum of the emitted gluon (see Ref.~\cite{gvw,ww00} for
similar formula in the GLV approach). Note that in Eqs.~(12)-(14)
in Ref.~\cite{azz} the transverse momentum of the emitted gluon is
replaced by $\vec\ell_T\rightarrow \vec\ell_T-\vec k_T$. Such replacement
must be kept in mind when one expands the hard part in the initial
transverse momentum $k_T$.

The above spectrum is proportional to the transverse momentum dependent
gluon distribution. One can interpret the corresponding hard part as the
partonic cross section of quark scattering with a {\it physical} gluon.
It vanishes $\widetilde H(\vec k_T)=0$ at $\vec k_T=0$, which is
quite different from the hard part $H_2^{--}(k_T)$
of quark and longitudinal gluon interaction in the higher-twist
approach in a covariant gauge. 
Therefore, making a
collinear expansion of this partonic cross section does NOT correspond
to the collinear expansion of the hard part of quark and gluon
interaction in the higher-twist approach in the covariant gauge.
In order to compare to
the results [Eq.~(\ref{crss0}) or (\ref{diff3})] in the collinear 
factorized formulation of higher-twist approach, one should 
integrate the above LCPI result over initial gluon's
transverse momentum while keep the transverse momentum $\vec\ell_T$
of the emitted gluon fixed.

One can similarly make a small $k_T$ expansion of the hard
part of the above LCPI result as in the higher-twist approach,
\begin{equation}
\widetilde H(\vec k_T)=4\pi\alpha_s^2
\frac{(\vec k_T \cdot\vec\ell_T)^2}{\ell_T^6}
\left[1-\cos\frac{\ell_T^2y_N^-}{2q^-z(1-z)}\right]
+{\cal O} (k_T^3) .
\end{equation}
After integrating over the initial gluon's transverse momentum
and defining the $k_T$-integrated gluon distribution function
\begin{equation}
xG_N(x)=\int dk_T^2 \frac{xdG_N(k_T^2,x)}{dk_T^2},
\end{equation}
one can obtain exactly the same gluon spectrum as in
Eq.~(\ref{diff3}) induced by hard-soft scattering in the
higher-twist approach. The effect of higher order terms in the
above expansion will be suppressed by powers
of $\langle k_T^2\rangle/\ell_T^2$ and
$\langle k_T^2\rangle \ell_T^2(R_A/q^-)^2$ as in the collinear
expansion in the higher-twist approach. Similarly,
this approximation is no longer valid for small values of
$\ell_T^2<\langle k_T^2\rangle$.

To study the effect of these higher order terms in the
$k_T$ expansion of the LCPI approach, one needs to know the
form of $k_T$-dependent gluon distribution
function $G_N(k_T^2,x)$. In the GLV approach \cite{Gyulassy:2000er},
a static potential model was used for quark medium interaction.
The induced gluon spectrum can be written as \cite{gvw,ww00}
\begin{eqnarray}
\frac{dN_{\rm GLV}}{dzd\ell_T^2}&=&\frac{C_A \alpha_s}{\pi^2}
\frac{1+(1-z)^2}{z}
\int dy_N^- \rho_A(y_N)
\int d^2k_T \frac{\sigma_{qN}\mu^2}{(k_T^2+\mu^2)^2}
\nonumber \\
&&\hspace{0.5in}\times
\frac{\vec k_T\cdot\vec\ell_T}{\ell_T^2(\vec k_T-\vec\ell_T)^2}
\left[1-\cos\frac{(\vec\ell_T-\vec k_T)^2y_N^-}{2q^-z(1-z)}\right],
\end{eqnarray}
where $\sigma_{qN}$ is the quark-parton scattering cross
section in the medium and $\mu^2$ is
the screening mass in the static potential.

If one make a similar small $k_T$ expansion of the hard part in the
above GLV result
\begin{eqnarray}
&&\frac{\vec k_T\cdot\vec\ell_T}{\ell_T^2(\vec k_T-\vec\ell_T)^2}
\left[1-\cos\frac{(\vec\ell_T-\vec k_T)^2y_N^-}{2q^-z(1-z)}\right]
\nonumber \\
&&=2\frac{(\vec k_T \cdot\vec\ell_T)^2}{\ell_T^6}
\left[1-\cos\frac{\ell_T^2y_N^-}{2q^-z(1-z)}\right]
+{\cal O} (k_T^3),
\end{eqnarray}
the induced gluon spectra in Eq.~(\ref{diff3}) from hard-soft scattering
in the higher-twist approach can be recovered if one identifies the
soft gluon distribution function as,
\begin{equation}
\frac{2\pi^2\alpha_s}{N_c} x_\mu G_N(x_\mu)
=\sigma_{qN} \mu^2 \log\frac{Q^2}{\mu^2}.
\label{glvgluon}
\end{equation}
The above equation can be rewritten as,
\begin{equation}
\frac{4\pi^2\alpha_s C_F}{N_c^2-1} x_\mu G_N(x_\mu)
=\int dk_T^2\frac{d\sigma_{qN}}{dk_T^2} k_T^2.
\end{equation}
which relates the soft gluon distribution of the medium and the average
transverse momentum weighted cross section in the higher-twist 
approach \cite{Baier:1996sk,jorge}.

Without the small $k_T$ approximation, one can make a variable 
change $\vec\ell_T^\prime= \vec\ell_T-\vec k_T$
in the integration over $\vec\ell_T^\prime$, and complete the 
integration over the initial gluon transverse momentum $\vec k_T$,
\begin{eqnarray}
\frac{dN_{\rm GLV}}{dz}&=&\frac{C_A\alpha_s}{2\pi}
\frac{1+(1-z)^2}{z}
\int dy_N^- \rho_A(y_N) \sigma_{qN}\mu^2
\nonumber \\
&&\hspace{0.5in}\times\int  
\frac{d\ell_T^{\prime 2}}{\ell_T^{\prime 2}(\ell_T^{\prime 2}+\mu^2)}
\left[1-\cos\frac{\ell_T^{\prime 2}y_N^-}{2q^-z(1-z)}\right].
\label{glv}
\end{eqnarray}
The above result can be compared to the induced gluon spectrum
in the higher-twist approach after regularization in $\ell_T^2$,
though one should note that $\ell_T^\prime$ after the variable
change in the above spectrum is no longer the true transverse
momentum of the radiated gluon, which can be integrated over to
obtain the transverse momentum integrated gluon spectrum $dN/dz$.
The above spectrum can be compared to the collinear
factorized form of the spectrum in Eq.~(\ref{diff3}), if one
identifies
\begin{equation}
\frac{2\pi^2\alpha_s}{N_c} x_\mu G_N(x_\mu)
=\sigma_{qN} \mu^2
\end{equation}
which differs from Eq.~(\ref{glvgluon}) in the small $k_T$ approximation
by a logarithmic factor
\begin{equation}
\mu^2\rightarrow \frac{1}{\sigma_{qN}}
\int dq_T^2\frac{d\sigma_{qN}}{dq_T^2} q_T^2\approx\mu^2\log\frac{Q^2}{\mu^2}.
\end{equation}
Therefore, one can consider the above logarithmic factor as the conseqeunce of
corrections beyond the small $k_T$ approximation within the static model of 
the GLV approach.

\section{Summary}

We have briefly reviewed in this paper the framework of twist
expansion for the calculation of higher-twist contributions to
semi-inclusive cross section of DIS off a nucleus target.
We also illustrated the simple partonic picture of the higher-twist
result for the induced gluon spectrum and demonstrated its
equivalence to the quark-gluon scattering cross section in
the collinear factorized framework. We
pointed out AZZ's misunderstanding of the collinear expansion
in this framework and their confusion over the procedure
and region of validity of the collinear approximation.
We showed that the small $k_T$ expansion in the LCPI
approach without the misconstrued variable change leads
to the same semi-inclusive gluon spectrum as in the higher-twist
approach. We showed that corrections to the
small $k_T$ (or collinear) approximation within
the static potential model in the GLV approach lead
to a logarithmic factor.

\section*{Acknowledgement}
The author would like to thank A. Majumder and B.-W. Zhang
for helpful discussions.
This work is supported  by the Director, Office of Energy
Research, Office of High Energy and Nuclear Physics, Division of
Nuclear Physics, of the U.S. Department of Energy under Contract No.
DE-AC02-05CH11231


\begin{thebibliography}{99}
\bibitem{note1} One should in principle include contributions from
interaction with transverse gluon fields. These contributions can be
combined with different terms in the collinear expansion of the
corresponding hard part to form the leading higher-twist contribution
of the cross section. See Ref.~\cite{lwz2} for detailed proof.
\end{thebibliography}
\end{document}